\documentclass[conference]{IEEEtran}
\usepackage[utf8]{inputenc}

\usepackage{todonotes}
\usepackage{tabu}
\usepackage{enumitem}
\usepackage[fleqn]{mathtools}
\usepackage{url}

\title{An Early Benchmark of  Quality of Experience Between HTTP/2 and HTTP/3 using Lighthouse}

\author{
  \IEEEauthorblockN{
    Darius Saif\IEEEauthorrefmark{1}, Chung-Horng Lung\IEEEauthorrefmark{2}, Ashraf Matrawy\IEEEauthorrefmark{3}
  }
  \IEEEauthorblockA{
    Carleton University, Department of Systems and Computer Engineering\\
    Email: {Dariussaif\IEEEauthorrefmark{1},Chlung\IEEEauthorrefmark{2},Amatrawy\IEEEauthorrefmark{3}}@sce.carleton.ca 
  }
}

\begin{document}
\maketitle

\begin{abstract}
Google's QUIC (GQUIC) is an emerging transport protocol designed to reduce HTTP latency. Deployed across its platforms and positioned as an alternative to TCP+TLS, GQUIC is feature rich: offering reliable data transmission and secure communication. It addresses TCP+TLS's (i) Head of Line Blocking (HoLB), (ii) excessive round-trip times on connection establishment, and (iii) entrenchment. Efforts by the IETF are in progress to standardize the next generation of HTTP's (HTTP/3, or H3) delivery, with their own variant of QUIC. While performance benchmarks have been conducted between GQUIC and HTTP/2-over-TCP (H2), few analyses, to our knowledge, have taken place between H2 and H3. In addition, past studies rely on Page Load Time as their main, if not only, metric. The purpose of this article is to benchmark the latest draft specification of H3 and dig into a user's Quality of Experience (QoE) by using Lighthouse: an open source (and metric diverse) auditing tool. Our findings show that, for one of H3's early implementations, H3 is mostly worse but achieves a higher average throughput.
\end{abstract}


\IEEEpeerreviewmaketitle

\section{Introduction}
QUIC is an emerging transport protocol which has been developed, and rolled out across services, by Google \cite{langleyQuicSpec}. Its features akin to TCP+TLS (such as loss and congestion control, security \cite{lychevSecurityPerformance}, Forward Error Correction (FEC) \cite{qian2018achieving}, and even multi-path \cite{de2017multipath}) position QUIC as an alternative to the former two. QUIC also brings advanced features like stream multiplexing to the table.

The primary motivation for QUIC is to reduce web page latency, thus bolstering a user's Quality of Experience (QoE) \cite{qiu2019experimental}. QUIC's major advantages over TCP+TLS are (i) eliminating Head-of-Line Blocking (HoLB) through stream multiplexing of independent resources, and (ii) fewer Round-Trip Times (RTTs) required on connection establishment, thanks to QUIC's cross-layer design. Google researchers have proposed a disruptive approach rather than extensions to TCP most notably because of TCP’s entrenchment in networks and Operating Systems (OS). Rather, QUIC is rapidly deployable, as it runs in user space.

\begin{figure}[!htb]
\centering
\includegraphics[width=3.4in]{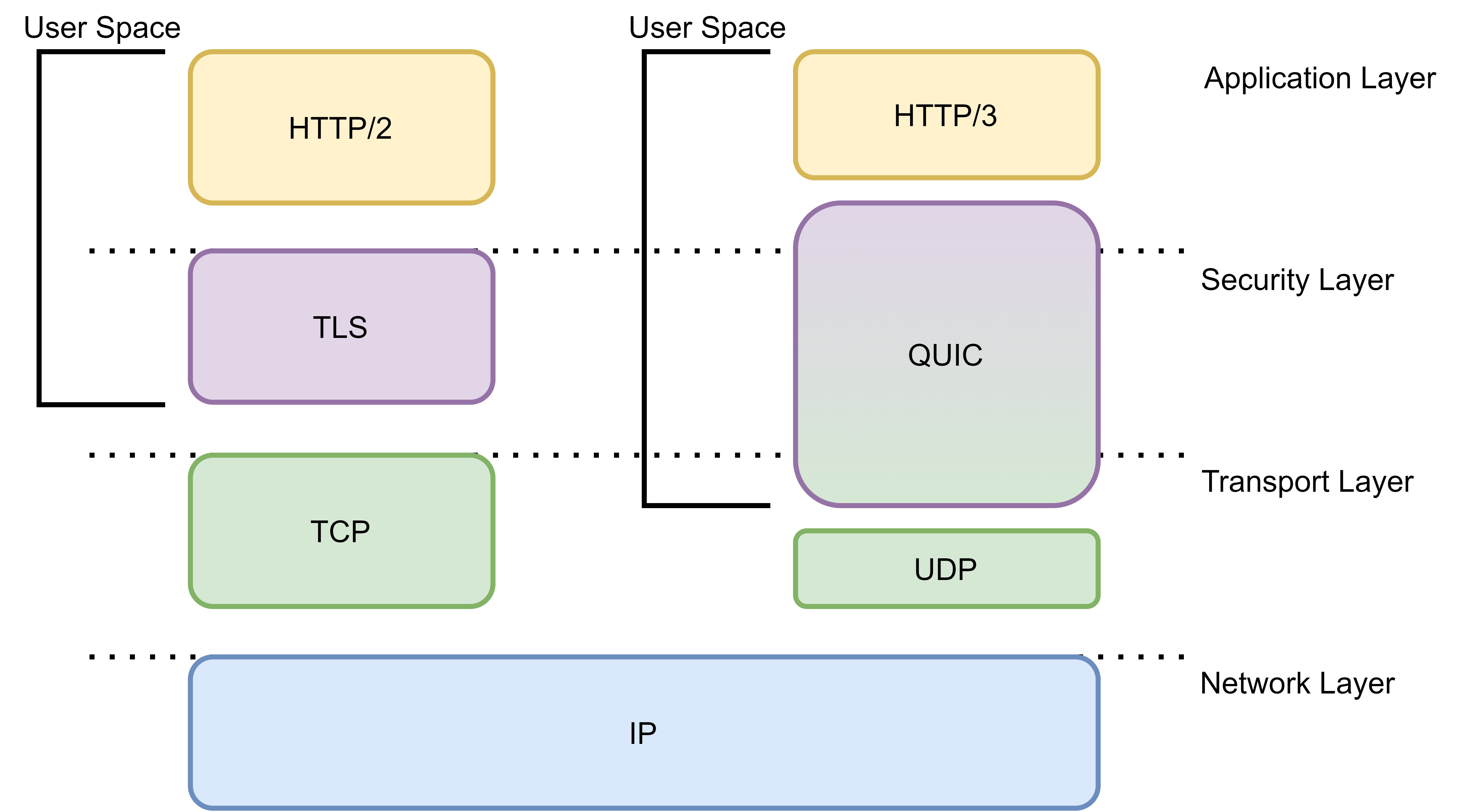}
\caption{Traditional HTTP/2 Stack vs. HTTP/3}
\label{stacks}
\end{figure}

The IETF has begun standardizing their own variant of QUIC. This transport has become the backbone of the next generation protocol HTTP/3 (H3) \cite{bishop2019hypertext}, whereby HTTP semantics are mapped over QUIC. A comparison of HTTP/3 and HTTP/2 stacks is illustrated in Figure \ref{stacks}. Given this, Google’s implementation is now commonly referred to as GQUIC.

Performance comparisons between (G)QUIC and TCP+TLS have primarily considered Page Load Time (PLT). This article's purpose is to extend upon those analyses from the standpoint of providing better visibility on QoE. As such, use of Lighthouse \cite{lighthouse} is proposed. It is an open source auditing tool which provides information-rich metrics and an aggregate performance score. Lighthouse is able to capture QoE features (like HoLB and prioritization) which a PLT analysis cannot.

This article provides four main points of contribution: (i) an early look at H3's performance, which has taken place in one other study \cite{marxresource}, to our knowledge, (ii) comparison against HTTP/2 (H2) over TCP+TLSv1.3, which is more competitive than TCP+TLSv1.2 in terms of connection establishment, (iii) a discussion on how the differences between GQUIC and IETF QUIC may affect their respective performance, and (iv) incorporating more metric diversity into test scenarios to better represent QoE implications, not widely considered before. 

Studies on GQUIC \cite{carlucci2015http,kakhkiRigorousEvaluation} found it to be more suitable than H2-over-TCP+TLSv1.2 in networks with high RTT. Our study between H3 (with IETF QUIC, hereby simply called QUIC) and H2-over-TCP+TLSv1.3 did not yield the same observation. Explanations to this are offered in this article and we invite others to reproduce, and confirm, the results. Our benchmarking was performed on Chrome Canary to an NGINX server with a custom CloudFlare patch to support H3.

The rest of this article is organized as follows: Section II surveys related works in this area. Details of the setup and metrics used are covered in Sections III and IV, respectively. Then, the benchmarking methodology and results are presented in Section V and VI. Discussion on the results and the article's conclusions are made in Sections VII and VIII. Finally, future works are presented in Section IX.

\section{Related Work}
Because of (G)QUIC's infancy, a number of server implementations, in addition to live traffic testing \cite{kakhkiRigorousEvaluation,biswalWebFaster,cookBetterForWhom,yu2017quic,megyesi2016quick,kharat2018quic}, have been considered in the literature.

Carlucci \textit{et al.} \cite{carlucci2015http} considered goodput (the average network receive rate minus retransmissions and FEC), channel utilization, loss ratio, and PLT in their analysis of GQUIC v21 and HTTP/1.1. Both used congestion control from \cite{ha2008cubic}. They found GQUIC had higher goodput in under-buffered networks, fared better in lossy networks, and reduced PLT. FEC, not enabled by default, noticeably worsened GQUIC's performance. 

Cook \textit{et al.} \cite{cookBetterForWhom} created a scriptable tool, \textit{Perfy}, to test PLT of HTTP/1.1, H2, or H2-over-QUIC pages in mobile and ADSL networks. Go-QUIC \cite{lucasClem} was used to power their server; hosting replicas of popular websites. They had found that QUIC fared better in mobile networks, but its gains were not as pronounced in more reliable settings.

Biswal \textit{et al.} \cite{biswalWebFaster} used Chromium's GQUIC v23 server. Unlike \cite{cookBetterForWhom}, their pages were engineered to be of certain sizes and numbers of Document Object Models (DOMs). They concluded that, as the size of objects on a page increased, GQUIC outperformed H2. Conversely, with more small objects per page, H2 fared better. This was noted as counter-intuitive due to GQUIC's theoretical edge by means of stream multiplexing.

Fairness, video QoE, and proxying were tackled by Kakhki \textit{et al.}'s \cite{kakhkiRigorousEvaluation} study on GQUIC versions up to v34. They modified GQUIC's code to tune parameters and also print debug traces, enabling root cause analysis. They found that GQUIC was unfair to TCP+TLSv1.2 and mostly outperformed it on desktop and mobile. When either variable network delays or large numbers of small objects were considered, GQUIC performed significantly worse than TCP+TLSv1.2.

A similar argument against PLT was made in Wolsing \textit{et al.}'s works \cite{wolsing2019performance,ruth2019perceiving}. In \cite{wolsing2019performance}, visual metrics (like those of Lighthouse) were used to shed light on each protocol's performance. TCP parameters were closely tuned to those of GQUIC v43 using a MahiMahi testbed. Comparisons took place for DSL and LTE. It was found that, in many settings, GQUIC was still superior, but by a much smaller margin than other works. In \cite{ruth2019perceiving}, a similar testbed and metrics were used, this time with human observers rating a page's QoE, augmenting the visual metrics. GQUIC performed better but users weren't able to distinguish either protocol.

\section{Experimental Setup}

\subsection{Server Side Setup}
A Ubuntu 18.04.4 (kernel 4.15.0-88) Virtual Machine (VM) in VirtualBox hosted the web server. It was allocated 4 processors and 6 GB of memory. Cloudflare's \textit{QUIC, HTTP/3, etc.} (QUICHE) project \cite{quiche} was leveraged (up to commit \textit{98757ca}) to provide H3 draft 27, and TLSv1.3, support to an NGINX v1.16 web server. This implementation of QUIC is written in Rust and BoringSSL for TLSv1.3. Let's Encrypt \cite{encrypt} was used to generate trusted certificates, as QUIC does not accept self-signed certificates.

H3 support was advertised to clients in the \textit{alt-svc} header for HTTP connections to the server. Both H3 and H2-over-TCP+TLS connections employed TLSv1.3 handshaking and used CUBIC congestion control. Out-of-the-box TCP tuning and parameterization were used.

Cloudflare notes that their H3 patch is not officially supported by NGINX. More importantly, the feature is marked as experimental and is subject to limitations. For example, at the time of writing, H3's 0-RTT connection establishment was not implemented \cite{0rttcf}. Use of OpenLiteSpeed as a web server for was also considered, which offered a similar support and performance disclaimer. NGINX was chosen due, in part, to our familiarity with the web server.

\subsection{Client Side Setup}
The Windows 10 machine hosting the VM was used as the client, shown in Figure \ref{netSet}. Baseline measurements of client to server were taken in \textit{iperf} and showed a rate of 2.53Mbps and ping time of 43ms. The client was loaded with Google Chrome Canary: a nightly built version of Chrome with various experimental features, including IETF H3 draft support. On startup, Canary can be instructed to support and negotiate H3 draft specification 27 with compliant servers by providing the flags \textit{--enable-quic}, \textit{--origin-to-force-quic-on=example.com:443} and \textit{--quic-version=h3-27}. Other browsers like Firefox Nightly and Safari also have experimental support for H3, not enabled by default.

\subsection{Network Impairments}
\begin{figure}[!htb]
\centering
\includegraphics[width=3.2in]{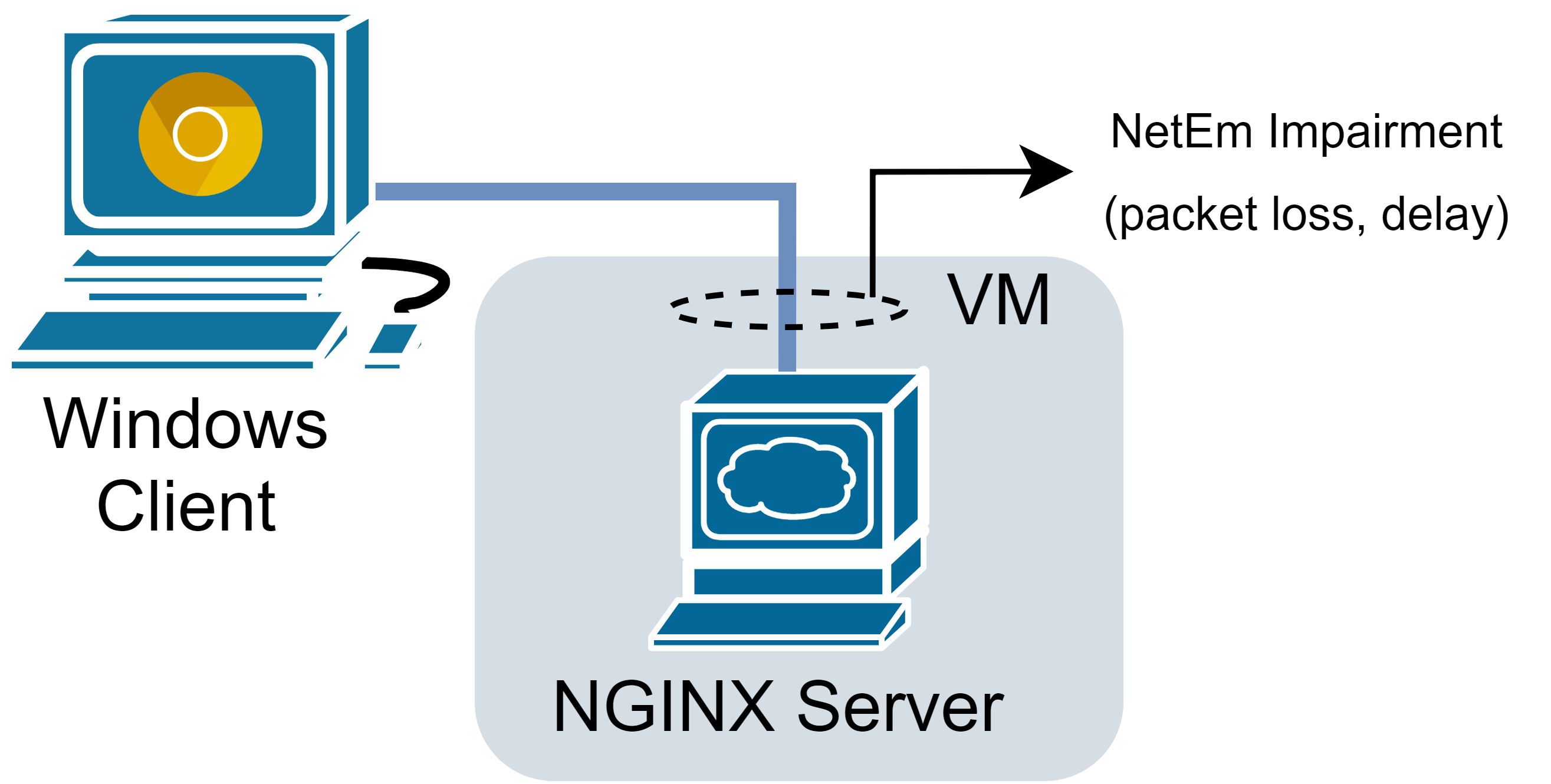}
\caption{Network Setup of Experimentation}
\label{netSet}
\end{figure}

NetEm \cite{hemminger2005network}, a standard Linux emulation tool, was used to control different network parameters, which was critical in benchmarking the respective protocols under various conditions. In this study, impairment rules were applied on outgoing packets on the server's network interface. Both packet loss and delay were considered, as shown in Figure \ref{netSet}. Other performance analyses \cite{qian2018achieving,kakhkiRigorousEvaluation,biswalWebFaster,cookBetterForWhom,megyesi2016quick} had also used NetEm to this effect.

\subsection{Web Content Served}

The web content used in every trial was designed to contain a mixture of content: CSS, JavaScript, text, and images, in order to resemble a realistic modern website. The web page's parameters of interest are presented in Table \ref{tab:pageParameters}.

{
\tabulinesep=1mm
\begin{table}[!htb]
  \centering
  \begin{tabu} to 0.47\textwidth {|X[]|X[]|}
     \hline
       Total DOMS&85 elements\\\hline
       Max DOM Depth&11 elements\\\hline
       Image Requests&12 (797KB)\\\hline
       Stylesheet Requests&2 (48KB)\\\hline
       Font Requests&1 (31KB)\\\hline
       Document Requests&1 (4KB)\\\hline
       Script Requests&1 (3KB)\\\hline
 \end{tabu}
  \smallskip
  \caption{Served Web Page Parameters}
  \label{tab:pageParameters}
\end{table}
}

HTTPArchive \cite{httpArchive} keeps track of statistics for millions of desktop and mobile URLs. In their reporting of early 2019, a desktop webpage had a median: (i) total payload of 1584KB, (ii) 658KB worth of image payload, and (iii) 398KB worth of JavaScript. In terms of payload and number of images, the served content is comparable to the reported median, although it is acknowledged that the served JavaScript payload is lower. 

\section{Performance Metrics}
\begin{figure*}[!t]
\centering
\includegraphics[width=7in]{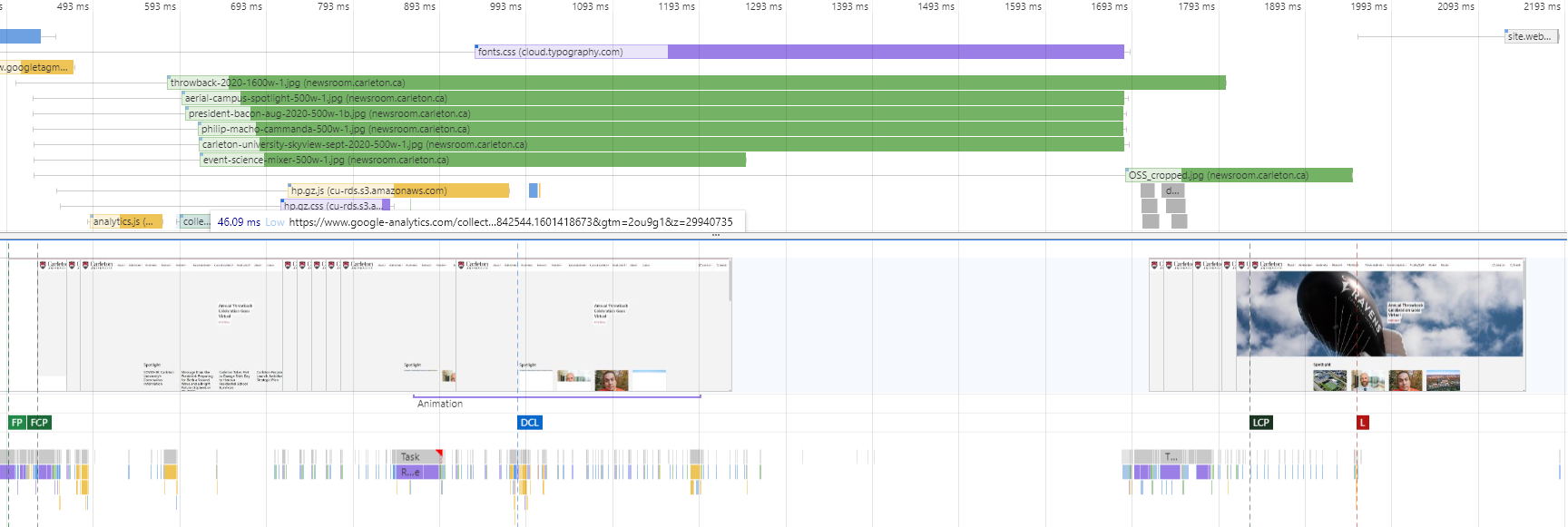}
\caption{Lighthouse Load Timeline Dashboard}
\label{lightDash}
\end{figure*}

Version 6.0.0 of Google's Lighthouse was leveraged as a tool for collecting QoE performance metrics. Lighthouse is an open source auditing tool included in Google Chrome's DevTools. It measures several characteristics of a web page (while the page loads) and groups them into 5 audit categories. The Performance category was of sole interest for this article. Lighthouse runs locally on a client machine and can be used on any website. The tool prepares a downloadable JSON report consisting of the recorded metric data and an interactive timeline of how the page rendered, shown in Figure \ref{lightDash} for an example web page.

Lighthouse's performance scoring scheme is comprised of three stages: first, raw values for the metrics are recorded. Then, individual metrics are ranked to a percentile, based on a log normal distribution of sample data from HTTPArchive. To limit outside factors in a web page's performance (network and device variation), a Lighthouse audit engages in CPU and network throttling to normalize sample data. Finally, the individual scores are combined according to a weighting system of each metric's impact on overall performance. The weights assigned to each metric are predetermined and are empirically derived by Lighthouse through heuristics.

The combined score, ranging from 0 (lowest) to 100 (highest), ultimately serves as a comprehensive indicator of the user's performance and QoE for a given page. Not only is the percentile ranking system for each metric publicly available, so too is the weighted metric combining scheme \cite{scoring}.

While Lighthouse measures a variety of performance metrics, only 6 are factored into the overall score in version 6.0.0. These metrics, and their weights, are presented in Table \ref{tab:lightMetric}.


{
\tabulinesep=1mm
\begin{table}[!htb]
  \centering
  \begin{tabu} to 0.47\textwidth {|X[1.4,r]|X[0.4]|X[4.7]|}
     \hline
      
      First Contentful Paint (FCP) &15\%&The time delta between first navigating to the web page and the browser rendering the very first DOM content.\\\hline
      
      Time to Interactive (TTI) &15\%&1. The FCP has completed\newline 2. Handlers are loaded for page elements\newline 3. The page responds to input within 50ms.\\\hline
      
      Speed Index (SI) &15\%&The time it takes for objects to be visibly displayed during page load.\\\hline
      
      Largest Contentful Paint (LCP) &25\%&The time it takes for the element on the page with the largest payload to have been completely rendered.\\\hline
      
      Total Blocking Time (TBT) &25\%&In the time between FCP and TTI, tasks taking longer than 50ms are summed into TBT. Timing starts after 50ms of task execution.\\\hline

      Cumulative Layout Shift (CLS) &5\%&Quantifies the page's stability as resources are loaded or DOMs are added. A higher score means more frequent layout shifts.\\\hline

 \end{tabu}
  \smallskip
  \caption{Lighthouse Performance Metrics}
  \label{tab:lightMetric}
\end{table}
}

The developers of Lighthouse, among other experts, maintain that PLT is subjective and loosely defined: arguing that page load does not occur at any \textit{single} instant but is rather a series of milestones. Factors including, but not limited to, HoLB and page resource prioritization have an impact on what content is populated when, and how interactive it is during load. These traits play in to the perceived responsiveness of a web page and are therefore directly tied in to the user's QoE.

The rich collection of metrics in Table \ref{tab:lightMetric} captures the full picture (request to load and everything in between) better than an analysis based purely on PLT, which skips over the user's experience during load. 
A similar observation is made in \cite{wolsing2019performance,ruth2019perceiving}, though metric combining was not covered in their work.

The meaning of raw data, particularly time deltas between two protocols, can be obscured without (i) a solid expectation on what objectively \textit{good} performance is, (ii) knowledge of the device(s) and network(s) under test, and (iii) specifics pertaining to the web content served: content type, payload, number of objects, etc. Lighthouse helps address these issues with its dashboard and percentile based scoring scheme.

\section{Procedure and Methodology}
Test cases were adapted from \cite{carlucci2015http,kakhkiRigorousEvaluation,biswalWebFaster,cookBetterForWhom}. Connections were generated through Lighthouse on Chrome Canary to the NGINX server. Only a single connection was made to the server at a given time. The protocol under test was toggled by starting Chrome Canary with or without the experimental flags. 

The browser's cache was cleared before performing each audit, eliminating the potential for either protocol's connection resumption to kick in. This helped ensure the benchmarking's fairness. A baseline with no NetEm impairment was captured for both protocols. Then, delay was incrementally introduced to create a higher RTT. At a fixed amount of delay, packet loss was then introduced and gradually increased. 

For each iteration, a total of 5 audits were performed and packet captures were taken in Wireshark. The raw Lighthouse metric data was averaged in order to deal with any variation. The largest difference in Lighthouse scoring between audits in each set was 6 points. The averaged raw metrics were then translated to an aggregate Lighthouse score, using the publicly available scoring calculator.

\section{Results}

\subsection{Baseline Measurement}

\begin{figure}[!htb]
\centering
\includegraphics[width=3.3in]{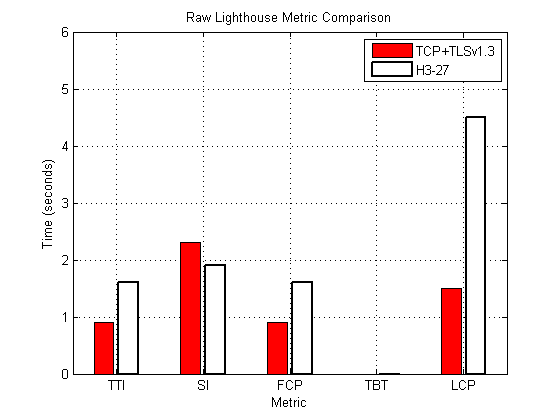}
\caption{Baseline Comparison of Raw Metrics}
\label{baseline}
\end{figure}

With no impairment from NetEm, a baseline was collected for each protocol. A histogram of the averaged raw metric data is provided in Figure \ref{baseline} -- the lower the value, the better. H3's SI beat its predecessor's but in terms of LCP, H3 fared decisively worse. The unitless aggregate Lighthouse scores for H3 and TCP+TLSv1.3's baseline were 65 and 87, respectively.

In studies related to GQUIC \cite{kakhkiRigorousEvaluation,cookBetterForWhom}, it had also been noted that, on a reliable connections (high-bandwidth, low loss, and low RTT conditions) with no impairment, TCP based delivery had an edge. It was shown in \cite{wangLinuxKernel} that this is because GQUIC introduced additional overhead by operating in user space rather than the kernel. The same holds true for H3.

CLS is not shown above as it's not measured in time -- it was 0.003 for both protocols. For all the conducted experiments, it was noted that the reported TTI and FCP were the same (that is, $TTI_{H3}$ = $FCP_{H3}$ and similarly for TCP+TLSv1.3), making TBT always 0ms. The line width of H3's bar graph makes TBT appear non-zero.

\subsection{Effects of Delay}

\begin{figure}[!htb]
\centering
\includegraphics[width=3.3in]{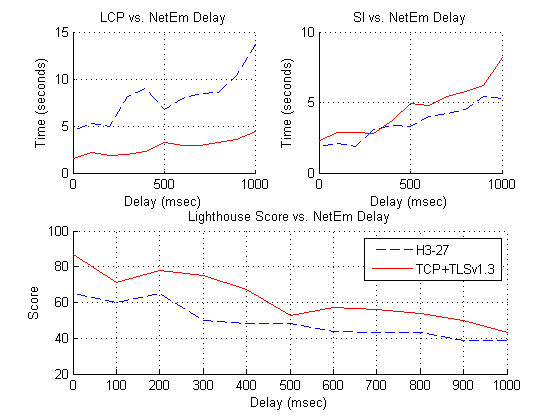}
\caption{Effects of Delay on Lighthouse Metrics \& Score}
\label{delay}
\end{figure}

For all proceeding plots, solid and dashed curves represent TCP+TLS1.3 and H3, respectively. Figure \ref{delay} shows raw LCP and SI values and the aggregate score. They were chosen to be highlighted since LCP is a highly weighted metric and H3 had a competitive SI. H3's aggregate score consistently trailed, and the largest score differentials (of 22 and 25) occurred with no impairment and at 300ms. Small spikes in the curves are due to the nature of the metrics: measuring both network conditions and the client's ability to paint content to the screen.

The performance gap became quite small approaching 1000ms of delay. TCP+TLSv1.3 still performed better in a trial with 2000ms delay. Although LCP was consistently much worse with H3, its SI remained competitive for the duration of testing. H3's TTI and FCP were consistently worse. 

\subsection{Effects of Packet Loss}

A fixed delay of 300ms and an increasing loss percentage were introduced with NetEm. This delay value was chosen since (i) it is widely accepted in the literature that QUIC fares better in unreliable conditions and (ii) buffer bloated mobile networks have been shown to have RTTs in the hundreds of milliseconds \cite{jiang2012tackling}.

\begin{figure}[!htb]
\centering
\includegraphics[width=3.3in]{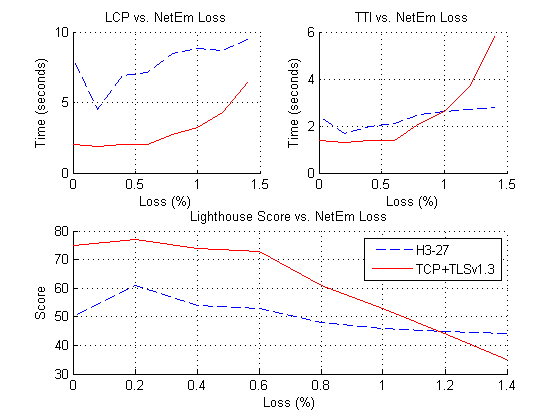}
\caption{Effects of Packet Loss on Lighthouse Metrics \& Score}
\label{loss}
\end{figure}

Figure \ref{loss} shows that, at higher loss rates, H3 overtook TCP+TLSv1.3. H3's aggregate score flattened out as more loss was introduced whereas TCP+TLSv1.3's curve decayed almost linearly. Beyond the setting of 1.4\% packet loss in NetEm, the results became quite unstable (in some cases Lighthouse was not able to properly complete its audit) and are hence not included.

Again, H3's LCP was much worse. However, H3's more stable TTI (and FCP) attributed to its higher scoring. The SI values between H3 and TCP+TLSv1.3 were very similar to one another in these trials. Packet captures in Wireshark showed that, with H3, almost twice as many packets were sent. The total aggregate bytes did not differ significantly however since the payload was about half.

\subsection{Throughput}

\begin{figure}[!htb]
\centering
\includegraphics[width=3.3in]{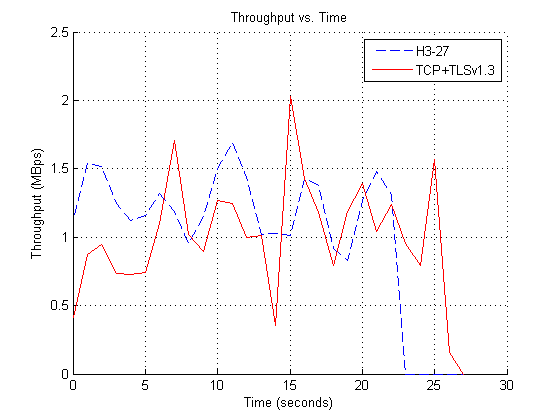}
\caption{25MB File Download Throughput}
\label{throughput}
\end{figure}

In this test, neither Lighthouse metrics nor NetEm impairment were used. A 25MB file download from the server was completed for both protocols, while capturing in Wireshark. Though the plots are superimposed, each file download occurred separately. The throughput results in Figure \ref{throughput} show that the file download finished 4 seconds faster in H3's favor.

Be that as it may, TCP+TLSv1.3 achieved a higher peak throughput (2.03MBps) than H3 (1.69MBps). The average throughput for H3 (1.24MBps) was more favorable than TCP+TLSv1.3 (1.03MBps). H3 produced 3\% more data on the wire, as it generated more packets than the stock TCP tuning. The fact that H3 finished faster made for an interesting comparison between the delay results presented in Figure \ref{delay} and the throughput results in Figure \ref{throughput}.

CloudFlare's CUBIC parameterization used more aggressive \textit{rwnd} and \textit{cwnd} values than that of TCP's stock tuning. These points certainly gave H3 somewhat of a head-start and kept its rate bounded between 1-1.5MBps. This advantage, however, didn't necessarily translate to a better Lighthouse score for the approximately 1MB web page. The larger buffers may have attributed to H3's generally better SI (weighted 15\%) scoring but its LCP (weighted 25\%) timings were more damning.

\section{Discussion}
Studies \cite{carlucci2015http,kakhkiRigorousEvaluation,biswalWebFaster,cookBetterForWhom,yu2017quic,megyesi2016quick,wangLinuxKernel} on (G)QUIC had identified scenarios which it had quite an edge over TCP+TLSv1.2. In comparing H3 to H2-over-TCP+TLSv1.3, the benefits were seldom apparent and rather marginal. Open source tools used in this work, and their software versions, have been detailed. Reproduction of these experiments can be accomplished with a similar testbed setup and webpage payload. We offer some early explanations as to why this may have been and invite further studies on H3:

\subsubsection{Lighthouse}
With stream multiplexing that addresses HoLB, it was expected that these metrics, and thus QoE, would favor H3 -- indeed SI, TTI, and FCP made for interesting outcomes. Under certain loss and high RTT conditions, these metrics fared better for QUIC. Alas, H3's consistently poorer performance, in terms of aggregate score, was because of its low LCP scoring. Lighthouse was not believed to have tipped the scales towards H2.

\subsubsection{Differences with GQUIC and QUIC}
Their state machines, source coding, and header framing contain innumerable differences. Notable examples include: (i) more fields in QUIC are encrypted, (ii) QUIC's method of header compression (QPACK \cite{ietf-quic-qpack-18}) is different from that of GQUIC's (HPACK), (iii) GQUIC uses a proprietary security scheme -- GQUIC Crypto \cite{langleyQuicSpec}, and (iv) GQUIC uses BBR \cite{cardwell2016bbr} congestion control.

Regarding point (iii), GQUIC's version 50 exists in two flavors: Q050 and T050 \cite{tquic}. The T050 designation is for the use of TLSv1.3 as the secure handshaking protocol and notes that it is somewhat of a halfway point between GQUIC and the IETF H3 drafts. Handshaking with TLSv1.3 is still under performance review by Google and T050 is said to only be in experimentation for 1\% of users. The other 99\% is made up from Q050 and older versions (like Q043 and Q046) which employ GQUIC Crypto \cite{tquic}.


\subsubsection{Limitations of Server Implementations}
It is stressed that implementations of H3 are made available for test purposes and do not claim to be suitable for production environments at this point. Chunks of the specification are either incomplete or subject to tuning and bug fixing. For example, at the time of experimentation, the CloudFlare implementation did not fully support 0-RTT connection establishment for QUIC.

H3 servers evolve quickly, just as the IETF drafts do. During the course of this experimentation, a number of updated drafts to H3 had been released, prompting a plethora of code churn in server implementations.

Recently, Cloudflare published a blog post \cite{cloudCompare} with initial testing of their own. It was found that for realistic pages, H3 was 1-4\% slower than H2. Although it is not clear if network impairment was considered, their results seem more or less consistent with the results presented in this article.

\subsubsection{H2-over-TCP+TLSv1.3}
In this article, H3 was benchmarked against H2-over-TCP+TLSv1.3, the latest version of TLS. Just like (G)QUIC, TCP+TLSv1.3 boasts a connection establishment of \textit{at most} 1-RTT (if TCP Fast Open \cite{cheng2014rfc} is used). Its predecessor, TCP+TLSv1.2, required 3-RTTs. Previous studies did not widely incorporate TLSv1.3 into their test environment, giving QUIC a performance edge of up to 3-RTTs. This made QUIC desirable by much higher margins in high RTT networks.

\section{Conclusions}
GQUIC is a low latency and entrenchment free alternative to TCP+TLS. Furthermore, its features and cross-layer design are able to address multiple TCP+TLS inefficiencies, like HoLB. Following deployment, and academic testing, of the protocol, the general consensus was that GQUIC was able to perform decisively better in environments with high RTT and/or packet loss as well as pages containing large objects. The IETF has modeled their transport protocol for the next generation of HTTP around these concepts.

The main, if not only, metric employed in most of the past works was PLT, which provides little insight into a user's QoE. Rather, this analysis leveraged Lighthouse for diverse metrics to depict various milestones throughout the page loading process. Until now, one other academic performance benchmark between H3 and TCP+TLSv1.3 has been presented \cite{marxresource}. It is acknowledged that the IETF specifications are merely drafts and that implementations were sparse and listed as experimental. Given that, our results showed that H3 mostly fared worse than its predecessor: H3 performed better under high loss and achieved a higher average throughput. Discussions and explanations as to why this may have been the case have also been provided.

\section{Future Works}

Due to the nature of QUIC -- whereby much of the transport level information is encrypted -- inspecting and debugging traffic becomes a particular area of focus \cite{marx2018towards}. Where traditional packet inspectors like Wireshark can come up with information-rich graphical interpretations of TCP flows, the same is not true for QUIC. Worse yet, the internal state machine of QUIC is not easily inferred from the state of traffic on the wire.

Marx \textit{et al.} \cite{marx2018towards} propose \textit{qlog} to this end: a incremental logging format for QUIC. Qlog is based on JSON and seeks to make problem root cause analysis for QUIC more accessible. More importantly, it has been integrated with some of the most prevalent QUIC client/server implementations \cite{marx2020debugging}.

As an extension of this work, it is planned to integrate qlog and its accompanying tools (\textit{qvis} and \textit{pcap2qlog}) into the test environment. The events captured during such experimentation will be leveraged in order to conduct root cause analysis into the relationship between the employed metrics and QUIC's congestion control, stream multiplexing, and packetization.

Other areas of future work include expanding test cases for 0-RTT connection establishment as well as breadth of web content served, such as progressive web apps.

\bibliographystyle{ieeetr}
\bibliography{references.bib}

\end{document}